\documentclass[11pt]{amsart}
\usepackage{amsfonts,eucal,amssymb}
\usepackage[all]{xy}
\begin{document}

\newcommand{\A}{{\mathcal A}}
\newcommand{\tA}{{\widetilde{A}}}
\newcommand{\tcA}{{\widetilde{\mathcal A}}}
\newcommand{\AC}{{\sf AC}}
\newcommand{\C}{{\mathbb C}}
\newcommand{\D}{{\mathbb D}}                                     
\newcommand{\R}{{\mathbb R}}
\newcommand{\uR}{{\underline{\mathbb R}}}
\newcommand{\ein}{{\mathcal E}}
\newcommand{\el}{{\mathbb L}}                          
\newcommand{\g}{{w}}
\newcommand{\bg}{{\overline{g}}}
\newcommand{\bbg}{{\bf g}}
\newcommand{\G}{{\mathbb G}}
\newcommand{\cG}{{\mathcal G}}
\newcommand{\tG}{{\widetilde{G}}}
\newcommand{\Gl}{{\rm Gl}}
\newcommand{\half}{{\textstyle{\frac{1}{2}}}}
\newcommand{\Oh}{{\rm O}}
\newcommand{\bPhi}{{\overline{\Phi}}}
\newcommand{\quarter}{{\textstyle{\frac{1}{4}}}}
\newcommand{\sgn}{{\rm sgn}}
\newcommand{\tv}{{\tilde{v}}}
\newcommand{\Tth}{{\textstyle{\frac{2}{3}}}}
\newcommand{\SU}{{\rm SU}}
\newcommand{\U}{{\rm U}}
\newcommand{\cS}{{\mathcal S}}
\newcommand{\tth}{{\textstyle{\frac{2}{3}}}}
\newcommand{\htt}{{\textstyle{\frac{3}{2}}}}
\newcommand{\tQ}{{\widetilde{Q}}}
\newcommand{\vol}{{\rm vol}}
\newcommand{\third}{\textstyle{\frac{1}{3}}}
\newcommand{\threehalf}{\textstyle{\frac{3}{2}}}
\newcommand{\Higgs}{{\rm Higgs}}
\newcommand{\Fermions}{{\rm Spinors}}
\newcommand{\Q}{{\mathbb Q}}

\title {On gauge theories of mass}
\author{Jack Morava}
\address{Department of Mathematics, Johns Hopkins University, Baltimore,
Maryland 21218}
\email{jack@math.jhu.edu}
\thanks{This research was supported by DARPA and the NSF}
\date {20 May 2010}

\begin{abstract}{The classical Einstein-Hilbert action in general relativity 
extends naturally to a blow-up (in the sense of algebraic geometry) of the 
usual space of pseudo-Riemannian metrics; this presents the metric 
tensor $g_{ik}$ as a kind of Goldstone boson associated to the real 
scalar field defined by its determinant. This seems to be quite compatible 
with the Higgs mechanism in the standard model of particle physics.}\end{abstract}

\maketitle

\noindent
{\bf 0.1} Gauge theories of mass can be traced back to Weyl's early work on conformal 
geometry [10 \S 1, 15 \S 16], in light of de Broglie's quantum-mechanical 
relation
\[
mc^2 = \hbar \nu
\]
between energy and frequency; since then, work in particle physics culminating 
in the standard model has made the subject predictive. This note argues that 
relations between physical mass and (Lorentz-Minkowski) geometry are best 
understood by thinking of the usual pseudometric tensor as the product of a 
`dimensionless' unimodular tensor (roughly $|\det g|^{-1/n}g_{ik}|$) with 
a suitable power of a density $\gamma \sim |\det g|^{-1/2n}$ (a section of 
a real line bundle which transforms like an inverse length). This enlarges 
the class of possible states by allowing $|\det g| \to 0$ or $\infty$, 
provided the conformal behavior is otherwise good. \bigskip

\noindent
Models of this sort have a long history in physics, but related recent developments 
in Riemannian geometry seem not to have had much impact there. In particular, 
Yamabe's nonlinear elliptic eigenvalue equation [11, 16]
\[
\Big[ -\Delta + \quarter \frac{n-2}{n-1} \; R(g) - \Lambda |u|^{4/(n-2)} \Big] \; u \; = \; 0 
\]
whose solutions $u > 0$ define conformal deformations $\bar{g} := u^{4/(n-2)}g$ of the
metric $g$ with scalar curvature 
\[
R(\bar{g}) \; = \; 4 \frac{n-1}{n-2} \Lambda \;=\; {\rm constant} 
\]
is (apart from the change from Riemannian to Minkowski signature) identical with 
a (super-renormalizable, if $n = 3, \; 4$ or 6) conformally invariant nonlinear wave 
equation with its own distinguished literature (cf. eg [3, 9, 13 \S 15.2]). \bigskip

\noindent
{\bf 0.2} Aside from an example in \S 4.2, there is little in the present account of this
material that is new, but the literature of these questions is old, broad, and perhaps 
confusing. My sense is that the subject is simpler than one might think; but there 
are nontrivial conceptual issues at stake, which are hard to present clearly. \bigskip

\noindent
I've tried to simplify things by keeping the language, and especially the physics, 
as classical as possible: in particular, though there is some discussion of the Higgs 
mechanism in \S 2.2, it occurs here only as a {\bf classical} field theory; delicate
quantum-mechanical issues involved in its renormalization are completely ignored.
However, there are some mathematical subtleties: the nonlinear eigenvalue problems
considered here may admit weak (ie moderately non-smooth) solutions, which correspond
to phase transitions in some physical models. The final example suggests
the possibility that a change in conformal structure in the interior Schwarzschild 
region might correspond to a change in the physical vacuum state there. \bigskip

\noindent
The paper is organized as follows: \S 1 summarizes some background, eg a 
blow-up of spaces of quadratic forms at 0, basic facts about densities on 
manifolds, and associated moduli spaces of geometric data. The next section 
uses this formalism to identify the Einstein-Hilbert action of classical 
relativity with a version of Yamabe's conformally invariant functional. 
\S 3 notes some consequences of the Sobolev embedding $L^2_1 \subset 
L^{2n/(n-2)}_0$ which seem to fit with old ideas from physics about Mach's 
principle, and \S 4 discusses some classical examples in this framework. \bigskip

\noindent
{\bf 0.3} I'd like to thank S. Agarwala and J. Baez for helpful, patient, and skeptical 
conversations in the early stages of this work, and Dan Christensen for unflappable 
help with confusing numerical matters toward the end. \bigskip

\begin{center}
{\bf \S 1 Desingularizing quadratic forms at 0}
\end{center} \bigskip

\noindent
{\bf 1.1} A nondegenerate symmetric bilinear form $q$ of signature $k$ on a 
finite-dimensional real vector space $V$ has a group $\Oh(q) \subset \Gl(V)$ 
of linear isomorphisms. If $n$ is the dimension of $V$, the homogeneous space 
$\Gl(V)/\Oh(q)$ is isomorphic to the space $Q^k_n(V)$ of nondegenerate 
quadratic forms of signature $k$ on $V$. This is an open cone, and its 
quotient
\[
\G_m\backslash \Gl(V)/\Oh(q) :\cong |Q|^k_n(V)
\]
by the center of $\Gl(V)$ is a space of `unimodular' quadratic forms of 
signature $k$, isomorphic to the (contractible) space of real $n \times n$ 
symmetric matrices whose eigenvalue configuration $\lambda_*$ consists of 
$n_\pm = \half(n \pm k)$ positive (resp. negative) real numbers satisfying
\[
(-1)^{(n-k)/2} \prod \lambda_* \; = \; 1 \;.
\]

\noindent
For any real $s$, let $\R(s)$ be the one-dimensional real representation of 
$\Gl_n(\R)$ defined by 
\[
\sgn (\det) \cdot |\det|^{s/n}: \Gl_n(\R) \to \G_m \;;
\]
it has a complex analog when $s \in \C$. \bigskip

\noindent
The quotient $\tQ^*_n$ of $Q^*_n \times \R(\half n - 1)$ defined by the 
action
\[
\rho \cdot (q, \; r) \mapsto (|\rho|^{4/(n-2)}q, \; \rho^{-1} r) 
\]
of $\rho \in \R^\times$ is a blowup
\[
q \mapsto [q,1] : Q^*_n \to \tQ^*_n \cong |Q|^*_n \times \R
\]
(the right-hand map sends $[q,r]$ to $(|\det q|^{-1/n} \cdot q, \; |\det 
q|^{(n-2)/4n}r)$) which replaces the cone of quadratic forms by the 
corresponding cylinder. \bigskip

\noindent
{\bf 1.2} For simplicity I'll assume $M$ compact closed, with principal frame 
bundle $\Gl(T_M)$, and I will write $\uR(s)$ for the real line bundle
\[
\R(s) \times_{\Gl_n(\R)} \Gl(T_M) \to M
\]
with fiber $\R(s)$. There is a canonical isomorphism
\[
\Lambda^nT^*_M \; \cong \; \uR(n)
\]
between $n$-forms on $M$ and sections of $\uR(n)$, and hence a canonical 
Lebesgue functional on the space $\Gamma \uR(n)$ of sections over $M$; 
sections of $\uR(0)$, on the other hand, are ordinary real-valued 
functions. When $n \geq s \geq 0$, let $\el(s)$ denote the real Banach space
of (equivalence classes of) sections of $\uR(s)$, completed with respect to 
the norm
\[
\|\phi\|_s \; := \; \Big[ \int_M |\phi|^{\otimes n/s}\Big]^{s/n}
\]
(or by the essential supremum when $s = 0$). H\"older's inequalities define a 
canonical isomorphism 
\[
\el(s)^* \cong \el(n-s)
\]
and continuous pairings
\[
\el(s) \otimes_\R \el(t) \to \el(s + t)
\]
when $0 \leq s,t,s + t \leq n$. $\Gamma \uR(s) \subset \el(s)$ will denote 
the subspace of smooth sections. \bigskip 

\noindent
{\bf 1.3} A pseudometric of signature $k$ on $M$ is a smooth section of the 
bundle 
\[
Q^k_n(M) := \Gl(T_M) \times_{\Gl_n(\R)} Q^k_n \to M \;;
\]
it defines a smooth section
\[
*_g^{s/n}(1) \;  := \; \sgn (\deg g) \cdot |\det g|^{s/2n} \in \el(s)
\]
of $\uR(s)$, as well as an isometry 
\[
\phi \mapsto *_g^{-s/n}\phi = f : \el(s) \to L^{n/s}(M,d\vol_g) \;.
\]
A pseudometric $g$ and a density $\phi \in \el(\half n - 1)$ define, at each 
point $x \in M$, an element $[g(x),\phi(x)] \in \tQ^*_n(T_{M,x})$ and thus a 
map
\[
\Gamma Q^*_n \times \Gamma \uR(\half n - 1) \to \Gamma \tQ^*_n
\]
which sends the pair $(g,\phi)$ to a section of the bundle of blown-up 
pseudometrics. However it is really the ray defined by $\phi$ that is 
significant, and it will be useful to assume that
\[
\phi \in \Gamma' \uR(\half n - 1) := \Gamma \uR(\half n - 1) - \{0\} \;.
\]
This constructs a kind of coarse moduli space of generalized pseudometrics, 
which has a finer analog: an element $u$ of the group $\Gamma \uR(0)^\times$ 
acts on $(g,\phi)$, sending it to $(|u|^{4/(n-2)}g, u^{-1}\phi)$, and the 
map to $\Gamma \tQ^*_n$ factors through the quotient of this action. \bigskip

\begin{center}
{\bf \S 2 A conformally invariant model} 
\end{center}

\begin{quotation}{``There is nothing in the world bigger than the tip of an autumn hair, and
Mount T'ai is little.'' \medskip

\noindent
Chuang Tzu, {\bf Discussion on making all things equal}, tr. Burton Watson}\end{quotation}
\bigskip

\noindent
{\bf 2.1.1 Proposition:} The diagram
\[
\xymatrix{
{\Gamma Q^*_n \times \Gamma' \uR(\half n - 1)} \ar[r] & {\Gamma Q^*_n 
\times_{\Gamma \uR(0)^\times} \Gamma' \uR(\half n - 1)} \ar[dr]^Y \ar[r] & 
\Gamma \tQ^*_n \ar@{.>}[d]^?\\
\Gamma Q^*_n \ar[u]_{1_{Q} \times \gamma} \ar[rr]^{\hbar^{-1} \ein}  & {} & 
\R}
\]
commutes; where
\[
\ein = \half \kappa^{-1} \int_M R(g) \; d\vol_g
\]
is the Einstein-Hilbert action functional (with $\kappa = 8\pi G$), 
\[
Y[g,\phi] \; = \; \half \int_M \Big[|df|^2_g +\quarter \frac{n-2}{n-1}\; 
R(g)f^2 \Big] \; d\vol_g
\]
is Yamabe's conformally invariant quadratic form\begin{footnote}{Here $Y$is 
normalized as if it were the Lagrangian in a Feynman measure of the form 
$\exp(-iL_{\rm matter}(\psi)/\hbar) D\psi$.}\end{footnote}(with $f$ as in \S 
1.3), and 
\[
\gamma \; := \; *_g^{(n-2)/2n}\big((\frac{n-2}{n-1} Gh)^{-1/2}\big) \;.
\] \bigskip

\noindent
{\bf Proof:} This is an absolute triviality (except for the assertion that 
$Y$ is conformally invariant, which is now classical [11, 16]). Note however that 
the dotted arrow is not asserted to exist. $\Box$ \bigskip

\noindent
{\bf 2.1.2} This reformulates the Einstein-Hilbert action as the Lagrangian 
for a conformally invariant physical theory involving a unimodular pseudometric
(ie a section of the bundle $Q^{-2}_4$) and a real spin-zero gauge 
field $\gamma$, defined locally by measurements of Newton's constant, with symmetry 
broken on the locus where $\gamma \to 0$ (or $\infty$, if we allow noncompact $M$). 
Away from this set, conformal invariance allows us to assume that (the function 
corresponding to the density) $\gamma$ is {\bf constant}, ie roughly the Planck 
frequency
\[
(\tth Gh)^{-1/2} \; \sim \; 90.7 \times 10^{35} \: {\rm MHz} \;. 
\]
This, after all, is what a gauge theory does; we understand $\gamma$ to be 
constant because it {\bf defines} the local mass scale. \bigskip

\noindent
At first sight $Y$ looks like the Lagrangian for a real scalar boson, moving 
in a potential field of the form $R(g)$ [9]; but requiring that $\phi$ not be 
identically zero can be interpreted as the introduction of a self-interaction 
term. The Sobolev embedding theorem says that (on a compact smooth 
$n$-manifold) the space $L^p_s$ of functions with $s$ derivatives in $L^p$ 
embeds in $L^q_t$ iff $t - n/q \leq s - n/p$: in particular, $L^2_1 \subset 
L^{2n/(n-2)}_0$ is just on the edge of continuity. Requiring that $\gamma$ 
have fixed norm as a $(\half n-1)$-density, ie that its Lebesgue 
$2n/(n-2)$-norm be finite, is equivalent to adding a Lagrange multiplier term 
of the form
\[
\Lambda (||\gamma||^{2n/(n-2)} - 1) \;,
\]
to the Lagrangian; which, when $n=4$, is equivalent to allowing the `dilaton' 
$\gamma$ a quartic (super-renormalizable) self-interaction. \bigskip

\noindent
{\bf 2.1.3} From this point of view, the `graviton' (ie, the field 
represented by the rank two symmetric tensor $g_{ik}$) is a Goldstone boson 
associated to $\gamma$: if we `decouple' the metric from its determinant by 
writing
\[
g_{ik} := \phi^{4/(n-2)}\bg_{ik}
\]
with
\[
\phi = (|\det g|^{1/2})^{(n-2)/2n}\in \Gamma \uR(\half n - 1) 
\]
(so $|\det \bg|^{1/2} = 1$), then Yamabe's equation
\[
\int *_gR(g) = \int \Big[\phi^2 R(\bg) + 4 \frac{n-1}{n-2}|d\phi|^2_\bg \Big] 
\; d^nx \;,
\]
is completely analogous to Goldstone's identity
\[
|d(e^{i\theta}\chi)|^2_g \; = \; \chi^2 |d\theta|^2_g \; + \; |d\chi|^2_g \;.
\]
The opposite interpretation -- that the dilaton is a Goldstone boson associated
to the metric -- is more usual in physics [12]. The interpretation here is that the boson
associated to the conformally invariant wave equation is more significant locally,
while (perturbations of) the Lorentz-Minkowski metric, though fundamental for 
geometry, become important only at quantum-mechanically vast distances. \bigskip

\noindent
In fact such issues go back to the earliest days of the subject. Weyl observed  
[15 \S 28] that the Einstein-Hilbert action can be written as a quadratic functional
\[
S(\dot{g}) := \int g^{ik}[\Gamma^s_{\; ts} \Gamma^t_{\; ik} - \Gamma^s_{\; 
it} \Gamma^t_{\;sk}] \; d\vol_g
\]
in the first derivatives of $g$, analogous to the left-hand side of Goldstone's identity.
\bigskip

\noindent
{\bf 2.2.1} It is well-known, but perhaps quite remarkable, that the standard model 
of particle physics is very close to conformally invariant; it is the usual coupling to 
gravitation which breaks the symmetry [4]. That model involves a principal bundle 
$P \to M$ with compact semisimple structure group 
\[
G \;= \; \SU(3) \times \SU(2) \times \U(1)
\]
as fiber; it postulates complex vector bundles associated to two 
representations of $G$, whose sections are called Higgs and fermion fields. 
I'll leave the details of these representations unspecified (see [5]): for 
our purposes the significant fact is that these are both bundles of 
$(\half n - 1)$-densities. \bigskip    

\noindent
The Lagrangian density of the standard model is a functional on a 
configuration space of fields $(A,\Phi,\Psi)$, equal to the sum of  
\bigskip

\noindent
$\; \bullet \;$ a Yang-Mills term $*_g|F_A|^2$ defined by the curvature $F_A$ 
of a connection one-form $A$ on $P$, \medskip

\noindent
$\; \bullet \;$ a Dirac term $\Psi^\dagger \cdot i {\displaystyle{\not} 
\partial}_A \Psi = L(A,\Psi)$, and  \medskip

\noindent
$\;\bullet \;$ a Higgs term $L(A,\Phi)$ of the form $|d_A\Phi|^2_g + P(\Phi)$, 
with the latter term a polynomial in $\Phi$, eg something like
\[
P(\Phi) \; = \; (|\Phi|^2 - \lambda^2 \phi^2)^2
\]
(with a {\bf dimensionless} coupling constant $\lambda$). \bigskip

\noindent
The Yang-Mills term is conformally invariant in dimension four, and if we 
rewrite the auxiliary fields 
\[
\Phi := *_g^{(n-2)/2n}\Phi_0, \; \Psi := *_g^{(n-2)/2n}\Psi_0
\]
in terms of fields $(\Phi_0,\Psi_0)$ of conformal weight zero, then the 
rescaling $g \mapsto u^{4/(n-2)}g$ sends 
\[
L(A,\Psi_0) \mapsto L(\tA,\Psi_0), \; L(A,\Phi_0) \mapsto L(\tA,\Phi_0)
\]
with $\tA := A + u^{-1}du$. \bigskip

\noindent
{\bf 2.2.2} Sections $\bbg$ of the bundle
\[
G^{\rm ad} \times_G P \to M
\]
(where $G^{\rm ad}$ is defined by the conjugation action of $G$ on itself) 
form a group $\cG(G)$ of gauge transformations, which act on the space $\A$ 
of connections on $P$ by $(\bbg,A) \mapsto A - d\bbg \cdot \bbg^{-1}$. The 
standard model Lagrangian is thus a function on the quotient
\[
(\Higgs \times \Fermions) \times_{\cG(G)} \A 
\]
(with $\bbg(\Phi,\Psi) = (\bbg \Phi,\bbg \Psi)$ on the Higgs and fermion 
fields).\bigskip

\noindent
The analogous symmetry group for the geometric sector is the group $\D$ of 
(orientation and spin-structure-preserving) diffeomorphisms of $M$. This acts 
on $\cG(G)$, and the moduli space of states for the standard model coupled to 
the usual version of general relativity is a bundle 
\[
(\Higgs \times \Fermions)\times_{\cG(G)} \A \; \to \; (\cdots) \; \to \; 
\Gamma Q^*_n/\D 
\;.
\]

\noindent
{\bf 2.2.3} The multiplicative group $\Gamma \R(0)^\times$ of 
nowhere-vanishing real-valued functions on $M$ [\S 1.3] can be equally well 
regarded as a group $\cG(\G_m)$ of gauge transformations associated to a 
principal bundle with structure group the noncompact torus $\G_m(\R) = 
\R^\times$. If we interpret the fields of the standard model as densities 
as above, then its moduli space of field configurations can be presented 
as the quotient
\[
(\Higgs \times \Fermions) \times_{\cG(\tG)} \tcA
\]     
with 
\[
\tcA := \A \times d\Omega^0(M), \; \cG(\tG) := \cG(G \times \G_m) \cong \cG(G) 
\times \cG(\G_m) 
\]
under the action 
\[
(\bbg,u) \cdot (A,\Psi,\Phi) := (A - u^{-1}du - d\bbg \cdot \bbg^{-1}, \; 
u^{-1} \bbg \Psi,\; u^{-1} \bbg \Phi) \;.
\]
The Lagrangian of the standard model coupled to gravitation then extends, as 
in \S 2.1, to a function on the quotient of the space
\[
\big((\Higgs \times \Fermions) \times \tcA \big) \times_{\cG(\G_m)} 
\big(\Gamma Q^*_n \times \Gamma'\uR(\half n - 1) \big) 
\]
of fields by the gauge group $\cG(\tG) \rtimes \D$. This defines a 
conformally invariant version of the standard model coupled to gravity,
whose solutions are classical off the singular locus $\gamma^{-1}(0)$.
\bigskip

\noindent
{\bf 2.2.4} This suggests several possible directions of extension, 
which I hope to discuss later: \medskip

\noindent
$\; \bullet \:$ the Connes-Chamseddine-Lott noncommutative version of the 
standard model [4]\medskip

\noindent
$\; \bullet \;$ supersymmetric (eg minimal) extensions of the standard model,\medskip

\noindent 
$\;\bullet \;$ Connes-Kreimer-Marcolli renormalization of the conformally 
invariant 4D $\phi^4$ model (which involves treating mass as a gauge field) 
[1]; and \bigskip

\noindent
$\; \bullet \;$ questions of classical analysis: I've evaded certain issues
by assuming the underlying space-time manifold to be compact, and by working
entirely with smooth sections. However, \S 4.2 below suggests the interest
of weak (eg $L^2_1$) solutions to these equations. \bigskip 
                                   
\begin{center}
{\bf \S 3 Inferences from scale invariance} 
\end{center} \bigskip

\begin{quotation}{`` To show the fly the way out of the fly-bottle \dots'' 
\medskip

Ludwig Wittengstein, {\bf Philosophical Investigations}}\end{quotation} \bigskip

\noindent
Invariance under rescaling by a {\bf constant} factor $\rho > 0$ may help 
illuminate some basic issues of physical interpretation:
\[
R(\rho^{4/(n-2)}g) = \rho^{-4/(n-2)}R(g) \;,
\]
so 
\[
\|R(g)\|_{L^{n/2}(g)} = \Big[ \int_M |R(g|^{n/2}\; d\vol_g \Big]^{2/n}
\]
is scale-invariant, i.e. a `pure number'. In view of the extensive literature 
concerned with anomalously large or small cosmological numbers, it seems 
remarkable that this $L^2$-norm (when $n=4$) seems to be well-behaved in the 
standard astrophysical models: the current bound for the cosmological constant 
is roughly
\[
|\Lambda| \leq 10^{-35} \; {\rm sec}^{-2} \;,
\]
while the universe is thought to be something like $4 \times 10^{17}$ seconds 
old, suggesting
that 
\[
\|R\|_{L^2} \; \sim \; O(1) 
\]
(consistent with the hypothesis, plausible on other grounds, that $R=0$). 
\bigskip

\noindent
More generally, H\"older's inequalities imply that for a smooth real-valued 
function $f$ on $M$
we have
\[
\|R(g)f^2\|_{L^1(g)} \leq \|R(g)\|_{L^{n/2}(g)} \cdot \|f\|^2_{L^{2n/(n-
2)}(g)} \;.
\]
The norm appearing on the right rescales like a length:
\[
\|f\|_{L^{2n/(n-2)}(\rho^{4/(n-2)}g)} = \rho \; \|f\|_{L^{2n/(n-2)}(g)} \;,
\]
which suggests regarding it as an estimate of the `radius' of $M$ (i.e. the 
$n$th root of its volume), measured in units defined by $f$. On the other 
hand, interpreting $f$ as the inverse Planck length suggests regarding 
$R(g)f^2$ as an analog of the stress-energy scalar 
\[
T \; \sim \; \kappa^{-1}R(g) \;.
\]
In fact
\[
\|R(g)f^2\|_{L^1(\rho^{4/(n-2)}g)} = \rho^2 \|R(g)f^2\|_{L^1(g)}
\]
scales like $\|T\|_{L^1(g)} \; \sim \;  {\rm Energy}^2 \cdot {\rm 
Hypervolume}$, cf. [8 \S 3.3]. The inequality
\[
\|R\|_{n/2} \;  \geq \; \|Rf^2\|_1 \cdot \|f\|^{-2}_{2n/(n-2)}
\]
can therefore be interpreted as bound of the form
\[
{\rm Const} \; \geq \; \frac{{\rm Mass}}{{\rm Radius}} \;;
\]
conceivably this lies behind the `numerical coincidences' which physicists 
have interpreted as evidence for some version [3] of Mach's principle. 
\bigskip

\begin{center}
{\bf \S 4 Two examples}
\end{center} \bigskip

\begin{quotation}{`{}``The further in you go, the bigger it gets,'' said Hannah Noon.' 
John Crowley, {\bf Little, Big}}\end{quotation} \bigskip

\noindent
I'll close with an attempt to show that this formalism is not completely 
without content. The first example below is quite widely known [cf. eg. 
[9 \S 7]], but the second is more speculative. \bigskip

\noindent
{\bf 4.1.1} The positively curved Friedman pseudometric on $\R^4 \cong {\bf 
F}_+ $ is defined
by
\[
g \; = \; \left[\begin{array}{cc}
             1             &           0 \\
             0    &   - R^2A^2 {\bf 1}
            \end{array}\right] \;,
\]
where $A = (1 + \quarter r^2)^{-1}$ and $r^2 = x_1^2 + x_2^2 + x_3^2$. The 
expansion factor\begin{footnote}{I'm following the notation of [7 p. 117], 
but with $R$ denoting the Robertson-Walker factor.}\end{footnote} $R(t) \; (t 
= x_0)$ satisfies the differential equation
\[
\Big(\frac{dR}{dt}\Big)^2 \; = \; \frac{R_0 - R}{R} \;,\; R(0) = 0 \;;
\]
it increases to a maximum $R_0$ at time $t_0$, and then decreases to zero. 
The conformally equivalent pseudometric
\[
R(t)^{-2}dt^2 - A^2 \; \Sigma_{1 \leq k \leq 3} \; dx^2_k
\]
on the stereographic completion $\R \times S^3$ is isometric to a product 
manifold with time parameter $d\tau = R(t)^{-1}dt$, i.e. such that 
\[
\Big(\frac{dR}{d\tau}\Big)^2 \; = \; (R_0 - R)R \;,
\]
which is satisfied by $R(\tau) = R_0 \sin^2 (\tau/2)$. In other words, the 
spherical Friedman model is conformally equivalent to the universal cover of 
the compact manifold $S^1 \times S^3$ with the standard product Lorentz 
metric, endowed with a dilaton field (proportional to $\sin^2(\tau/2)$) 
which vanishes smoothly at singularities (big bangs) recurring with period 
$2\pi$. [This corresponds to identifying the upper and lower edges in the 
Penrose diagram displayed in Fig. 21(ii) of [8 \S 5.3].] \bigskip

\noindent
{\bf 4.1.2} The (hyper)volume of this model (i.e. of one cycle of the closed 
Friedman universe) seems not to be well known: it equals
\[
\int_{{\bf F}_+} *_g 1 \; = \; 2 \cdot {\rm Vol}(S^3) \cdot \int_0^{t_0} 
R(t)^3 dt \;.
\]
If $\g := R/R_0$ then
\[
R_0 \g' \;  =  \; (\g^{-1} - 1)^{-1/2} \;,
\]
so 
\[
dt \; = \; R_0(1 - \g)^{-1/2} \g^{-1/2} \; d\g
\]
and hence
\[
\int_0^{t_0}R(t)^3 dt = R_0^4 \int_0^1 \g^{5/2}(1 - \g)^{-1/2} \; d\g \;.
\]
Substituting $\g = \sin^2 \theta$ yields $2 \int_0^{\pi/2} \sin^6 \theta \; 
d\theta$ for the right-hand integral; this equals
\[
\frac{1}{96} \big[ 60\theta - 45 \sin 2\theta + 9 \sin 4\theta - \sin 6\theta 
\big]|^{\pi/2}_0 = \frac{5}{16} \pi \;,
\]
yielding (tip o'the hat to Archimedes) the value
\[
2 \cdot 2\pi^2 \cdot \textstyle{\frac{5}{16}}\pi \cdot R_0^4 \; = \; 
\textstyle{\frac{5}{4}} \pi^3 \cdot R_0^4
\]
for the volume of one Friedman $\AE$on. \bigskip

\noindent
{\bf 4.2.1} The Schwarzschild metric 
\[
d\tau^{2} = (1 - 2mr^{-1})dt^{2} - (1 - 2mr^{-1})^{-1}dr^{2} - 
r^{2}d\sigma^{2}
\]
is usually defined on the spacetime manifold $\R \times \R^{3} - {0}$;
\[
d\sigma^{2} = d\phi^{2} + \sin^{2} \phi d\theta^{2}
\]
is the standard metric on the two-sphere. More generally, the expression 
\[
d\tau^{2} = q(r)r^{-2}dt^{2} - q(r)^{-1}r^{2}dr^{2} - r^{2}d\sigma^{2}
\]
with 
\[
q(r) = r^{2} - 2mr + e^{2} = (r - m - D)(r - m + D)
\]
defines the Reissner-Nordstrom metric; the discriminant $D^{2} = m^{2} - 
e^{2}$ will be assumed positive here. Both of these examples have vanishing 
scalar curvature, but if $e > 0$ Reissner-Nordstrom space is not Ricci-flat.
\bigskip

\noindent
In null coordinates $v,w$ such that $t = \half (v + w)$ and 
\[
\half (v - w) = \int q(r)^{-1}r^{2}dr 
\]
the expression above defines the pseudometric 
\[
d\tau^{2} = qr^{-2}dv dw - r^{2} d\sigma^{2}
\]
with associated volume element
\[
|g|^{\half}dv \wedge dw \wedge d\theta \wedge d\phi = \half q \Omega dv \wedge dw 
= \Omega r^2 dr \wedge dt \;,
\]
where $\Omega = \sin \phi \; d\theta \wedge d\phi$ is the volume element for the 
two-sphere. The topology of the maximal analytic extension of such a 
pseudometric is complicated; see, for example, [8 \S 5.5]. Here we will be 
concerned mostly with the region of type II, in the terminology used there; 
this corresponds to the condition $r \in (m - D,m + D).$ 
\bigskip

\noindent
{\bf 4.2.2} In these coordinates the Laplace-Beltrami operator 
\[
\Delta f = |g|^{-\half} [ |g|^{\half} g^{ik}f_{,k} ]_{,i}
\]
takes the form 
\[
\Delta f = 2q^{-1}[(r^{2}f_{,v})_{,w} + (r^{2}f_{,w})_{,v}]
\]
for a function $f  = f(v,w)$ independent of $\theta$ and $\phi$. If $f$ is 
independent of $t$ as well, and we use primes to denote differentiation with 
respect to $r$, then $v' = q^{-1}r^{2}$ and 
\[
\Delta f = 2q^{-1}(r^{2}f'/q^{-1}r^{2})'/q^{-1}r^{2} = 2r^{-2}(qf')' \;.
\] 
There is thus a two-dimensional family of elementary harmonic functions $u = 
u(r)$ in the interior Schwarzschild region, characterized by the condition 
$u' = kq^{-1}$ for some constant $k$. Ignoring the time parameter, these 
functions are all of Lebesgue class $L^{4}$, but their derivatives are only 
locally of Lebesgue class $L^{2}$. Since $u'' = - kq^{-2}q'$, any such solution 
has a point of inflection at $r = m$.  The function defined by 
\[
U(r) = 1 + \frac {D} {m} \log \vert \frac {r-m-D}{r-m+D} \vert 
\]
if $r \in (m-D,m)$, and $U(r) = 1$ if $r \geq m$, is particularly 
interesting. It is continuous, but not differentiable, at $r = m$; its 
derivative is 
\[
U' = 2m^{-1}D^{2}q^{-1} \eta \;,
\]
where $\eta$ denotes a unit step function at $r=m$, so 
\[
\Delta U = 4m^{-3}D^{2} \delta
\]
with $\delta$ a Dirac delta-function at $r=m$. The piecewise-differentiable 
tensor $\bar g = U^{2}g$ thus has scalar curvature 
\[
\bar R = - 24m^{-3}D^{2}\delta
\]
which vanishes almost everywhere [14].\bigskip

\noindent
{\bf 4.2.3} The conformally deformed tensor $\bar g$ will not be Ricci-flat, 
even in the Schwarzschild case; a straightforward calculation [2 \S 6.3] 
shows that for a general conformal deformation $\bar g := u^{4/n-2}g$, 
\[
\bar R^{i}_{k} := R^{i}_{k}(\bar g) = u^{-4/n-2}R^{i}_{k}(g) + u^{-2n/n-
2}P^{i}_{k}
\] 
with 
\[
P^{i}_{k} = -2u[\nabla^{i}_{k}u + (n-2)^{-1} \delta^{i}_{k} \Delta u] + 2(n-
2)^{-1} [n \nabla^{i}u \nabla_{k}u - \delta^{i}_{k} |\nabla u|^{2} ] 
\]
where $\nabla_{i}$ signifies covariant differentiation [which on scalars is 
to be interpreted as ordinary differentiation]. If we assume that $n = 4$ and 
$u' = kq^{-1}$ as above, then the resulting tensor is diagonal in 
Schwarzschild coordinates; its entries have invariant significance,
as the eigenvalues of $P^{i}_{k}$ considered as an endomorphism of the 
tangent space. Displaying these diagonal components as vectors, we find that 
$P^i_k$ equals 
\[
kr^{-2}q^{-1}q'u[1,-1,0,0] + 2kr^{-3}u[-1,-1,1,1] + k^{2}r^{-2}q^{-1}[1,-
3,1,1] \;;
\] 
if $u' = kq^{-1} \eta$ is cut off at $r=m$, there is an additional term of 
the form $-kr^{-2} \delta [0,+1,0,0]$. Assuming $u = 1$ at its inflection 
point, $k = 2m^{-1}D^{2}$ is the unique value for which the determinant of 
$P^{i}_{k}$ vanishes at $r=m$; this characterizes the harmonic function $U$. 
\bigskip

\noindent
It is similarly straightforward to show that (as in the Schwarzschild case), 
the equation of a radial timelike geodesic in the metric $\bar g = U^{2}g$ 
becomes 
\[
U^{4} \dot r^{2} = b^{2} + (2mr^{-1} - 1)U^{2} \;, 
\]
where the dot denotes differentiation with respect to proper time, and $b$ is 
a constant of integration. When $r$ is small, 
\[
d\tau \sim (2m)^{-1/2} r^{1/2} U \;dr
\]
is integrable; as in the classical case, such a geodesic reaches the origin 
in finite proper time. \bigskip

\noindent
{\bf 4.2.4} This suggests the interest of solutions of the equation
\[
\Delta u \; + \; \Lambda u^3 \; = \; 0
\]
with $\Lambda \neq 0$. Taking $e=0$ and $m = \half$ for simplicity, and assuming as above that
$u$
depends only on the radial coordinate, this becomes
\[
2r^{-2}(r(r-1)u')' \; + \; \Lambda u^3 \; = \; 0 \;,
\]
which bears some (superficial?) resemblance to the Lane-Emden equation of astrophysics 
[6 Ch. IV].\bigskip

\noindent
By Cauchy-Kowalevskaya, the equation above has a unique solution analytic near 0 of the form
\[
3(2\Lambda)^{-1/2}v(r)  := 3(2\Lambda)^{-1/2}r^{-3/2} (1 + \sum_{k > 0} w_k r^k) \;,
\]
where
\[
(r(r-1)v')' \; + \; \textstyle{\frac{9}{4}}r^2 v^3 = 0 \;.
\]
In terms of $v(r) = r^{-3/2}w(r)$, the equation above becomes
\[
4r^2(r - 1) w''  + 4r(2 - r)w' + 3(r - 3)w + 9w^3 \; = \; 0 \;,
\]
which is satisfied by 
\[
w  \; = \; 1 - \textstyle{\frac{3}{26}} r - \textstyle{\frac{165}{26^2}} r^2 + \dots 
\]
Numerical computations (many thanks to S. Agarwala and Dan Christensen for invaluable help,
including infinitely many corrections) suggest this series has radius of convergence {\bf one},
and that $w$ is nonvanishing in the interval $[0,1)$. \bigskip

\noindent
The (orientation-reversing) monotonic change of variables $t = \log|r^{-1}-1|$ maps $(0,1)$ to 
$(-\infty,\infty)$. Rewriting the equation above as 
\[
v''(t)\; = \; \textstyle{\frac{9}{4}} \; \frac{e^t}{(1+e^t)^4} \; v(t)^3
\]
suggests that $v''(t) > 0$ for all $t \in \R$, and hence that the graph of $v$ is concave
upwards. Since $v(r) \to \infty$ as $r,s \to 0$, this implies the existence of a unique
critical point $v'(\rho) = 0$ with $\rho \in (0,1)$. Numerical calculations suggest
\[
\tv(r) \;=\; r^{-3/2}(1-r)^{-1/2}(1 - \Tth r)
\]
as a reasonable approximation to $v$ away from $r=1$; it has a unique critical point at 
\[
\tilde{\rho} \;=\; \quarter (7 - \surd 13) \; \sim \; .85\dots \;.
\]
Reasoning as in \S 4.2.2, this suggests that the function 
\[
V(r) = v(\rho)^{-1}v(r) \; {\rm if} \; r \in (0,\rho), \; = 1\; {\rm otherwise},
\]
defines a conformal deformation of the interior Schwarzschild metric with a 
second-order phase transition at $r = \rho$, with 
\[
\overline{R} = 6 \Lambda = 108m^2\rho^{-3}w(\rho)^2
\]
when $r < \rho$, such that 
\[
d\overline{\tau} \sim (2m)^{-1/2} \frac{\rho^{3/2}}{w(\rho)} \; w(r) \cdot r^{-1}dr
\]
has a logarithmic pole at $r=0$, defining a complete metric which puts the singularity
at infinity, possibly corresponding to an interesting new ground state for the interior 
Higgs field. The Penrose diagram for the associated conformally deformed metric 
glues together the vertical edges of Figure 25 in [8 \S 5.5], identifying horizontal 
pairs of parallel type I regions. \bigskip

\noindent
{\bf 4.2.5} In terms of the coordinate $t$, the equation above has an asymptotic solution
\[
v(t) \; \sim \; \sum_{k \geq 0} v_k t^{-k} \in \AC[[t^{-1}]]
\]
with coefficients in the differential Frech\'et algebra $\AC$ of smooth functions on $\R$ 
with rapidly decreasing (Schwartz class) derivatives. The function
\[
v_0(t) \;= \; \third[1 + 8(1+e^{-t})^{-2}]^{1/2} \; (= \third[1 + 8(1 - r)^2]^{1/2}) \in
\AC
\]
is an example: $v_0 \to 1$ as $t \to \infty, \; \third$ as $t \to -\infty$. If we define
\[
v(r) \; := \; r^{-3/2}(1 - r)^{-1/2} x(r) \;= \; 4e^t \cosh^2(\half t) \; x(t)
\]
then the equation for $v$ can be rewritten as a Duffing equation
\[
x''  \;+\; \delta_1 x' \;+\; \delta_0^2 x \; = \; \textstyle{\frac{9}{4}} x^3
\]
with
\[
\delta_1 = \frac{3e^t + 2 - e^{-t}}{e^t + 2 + e^{-t}}, \; \delta_0^2 = \frac{9e^t  + 2 + 
e^{-t}}{4(e^t + 2 + e^{-t})} \; \in \AC \;.
\]
This has $x_0 = \tth \delta_0$ (corresponding to $v_0$) as a kind of asymptotically stationary
approximate solution. \bigskip

\noindent
To improve the approximation, let $x = x_0 y$ (note that $\delta_0$ is invertible in $\AC$). The linear
operator
\[
L := \delta_0^{-3}(\partial + \delta_1)\partial \delta_0 = \delta_0^{-3}(\delta_0\partial^2 +
\epsilon_1 \partial + \epsilon_0)
\]
has coefficients
\[
\epsilon_1 = 2\delta_0' + \delta_1 \delta_0, \; \epsilon_0 = \delta_0'' + \delta_1 \delta'_0
\]
in the Schwartz class $\cS$. It thus extends to define a map from the differential algebra
$\AC[[t^{-1}]]$ to $\cS[[t^{-1}]]$. \bigskip

\noindent
Suppose now that $y(n) = \sum_{n \geq k \geq 0} y_k t^{-k} \in \AC[[t^{-1}]]$ has been
constructed, such that
\[
F(y(n)) \;:=\; Ly(n) + y(n) - y(n)^3 \in t^{-n-1}\AC[[t^{-1}]] \;;
\]
we can start an induction with $y_0 =1$, since $E_1 = t(\delta_0'' + \delta_1\delta_0') \in \cS$.
Then
\[
F(y(n) + y_{n+1}t^{-n-1}) \;\equiv\; F(y(n)) + F'(y(n))\cdot y_{n+1}t^{-n-1} \; {\rm mod}\; t^{-
n-2}\AC[[t^{-1}]]
\]
with $F(y(n)) \equiv E_{n+1} t^{-n-1}$ modulo higher powers of $t^{-1}$, for some
asymptotically constant error term $E_{n+1}$. The coefficient of $t^{-n-1}$ in the term
on the right above then simplifies to
\[
E_{n+1} + (L - 2)y_{n+1}
\]
and we can take $y_{n+1} = (2 - L)^{-1}E_{n+1}$.

\bibliographystyle{amsplain}

\end{document}